\documentclass{iaus}
\usepackage{graphicx}

\title[The third body in the eclipsing binary AV~CMi] 
{The third body in the eclipsing binary AV~CMi:
Hot Jupiter or brown dwarf?}

\author[Alexios Liakos, Dimitris Mislis and Panagiotis Niarchos]   
{Alexios Liakos$^1$,
\ Dimitris Mislis$^2$
\and Panagiotis Niarchos$^1$}

\affiliation{$^1$Department of Astrophysics, Astronomy and Mechanics, University of Athens, Athens, Hellas \\
email: {\tt alliakos@phys.uoa.gr,~}{\tt pniarcho@phys.uoa.gr } \\[\affilskip]
$^2$Institute of Astronomy, Madingley Road, Cambridge CB3 0HA, UK \\
email: {\tt misldim@ast.cam.ac.uk}}

\pubyear{2011}
\volume{282}  
\pagerange{111--112}
\jname{From Interacting Binaries to Exoplanets: Essential Modeling Tools}
\editors{A.C. Editor, B.D. Editor \& C.E. Editor, eds.}
\begin{document}

\maketitle

\begin{abstract}
New transit light curves of the third body in the system AV CMi have been obtained. The eclipsing pair's light curves were re-analysed with the W-D code and new absolute elements were derived for the two components. Moreover the new light curves (together with those given by \cite[Liakos \& Niarchos 2010]{L10}) of the third body transiting one of the components were analysed with the Photometric Software for Transits (PhoS-T). The results from both analyses are combined with the aim to study the nature of the third component.

\end{abstract}

\firstsection 
\section{Observations and analyses}

The new transit observations were obtained with a 40-cm Cassegrain telescope equipped with the CCD camera ST-10XME and by using only the I-filter, in order to achieve a better time resolution and higher signal-to-noise ratio. 11 new transit light curves were obtained in the years 2009-2011 increasing the total number to 18.


The light curves (hereafter LCs) of AV CMi were re-analysed with the PHOEBE v.0.29d software (\cite[Pr\v{s}a \& Zwitter 2005]{PZ05}) (for method details see \cite[Liakos \& Niarchos 2010]{L10}). The solution presented herein is based on a different assumed spectral type of the system (according to its B-V index (0.14-0.2)) as given in many catalogues (e.g. NOMAD, NPM2, ASCC-2.5 V3). Moreover, the absolute parameters of the eclipsing components were derived in order to check their evolutionary status and they were used for the calculation of some of the third body's characteristics as described below.

\begin{figure}[t]
\begin{center}
\begin{tabular}{cccc}
\includegraphics[width=3.1cm]{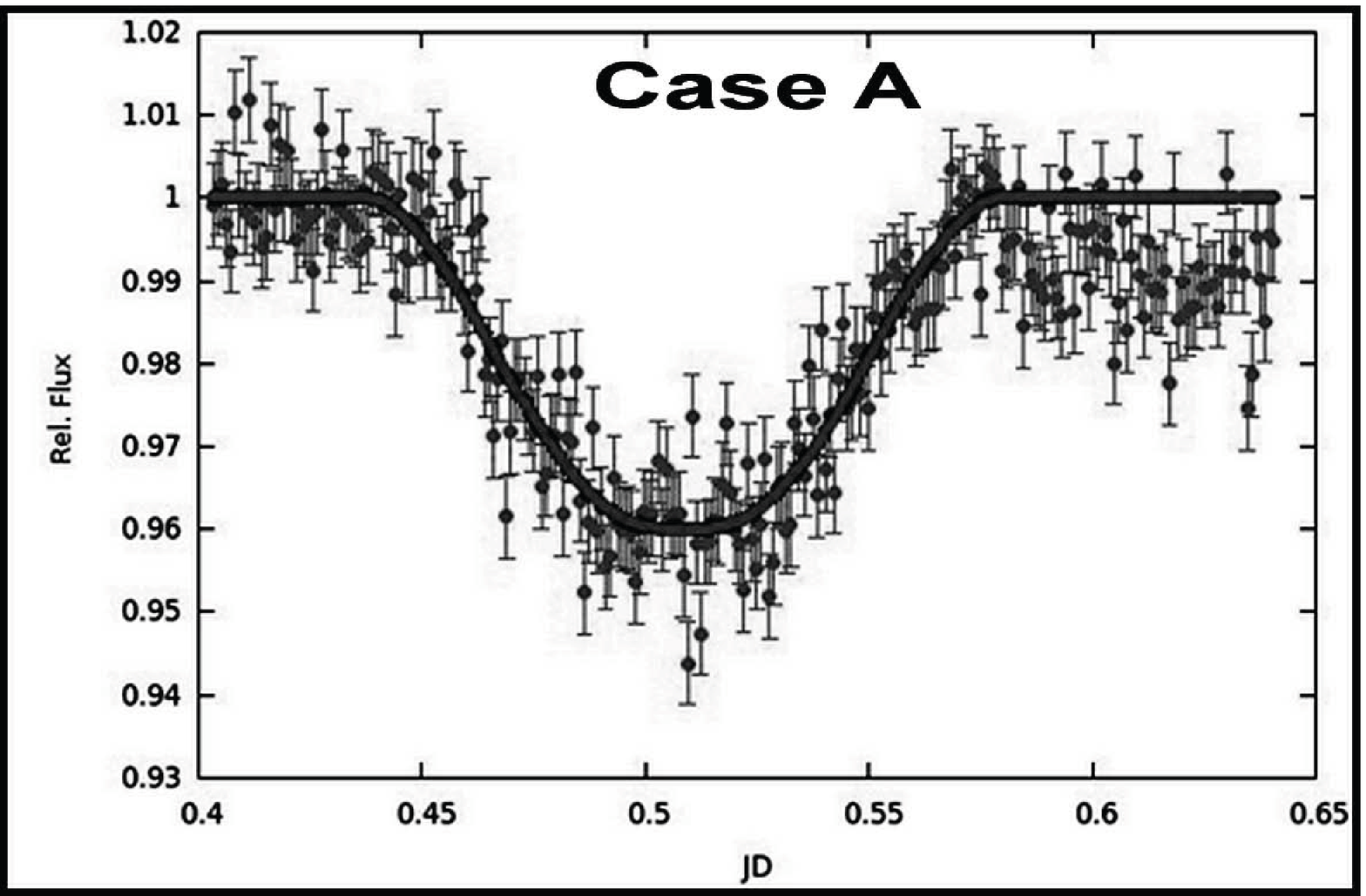}&\includegraphics[width=3.0cm]{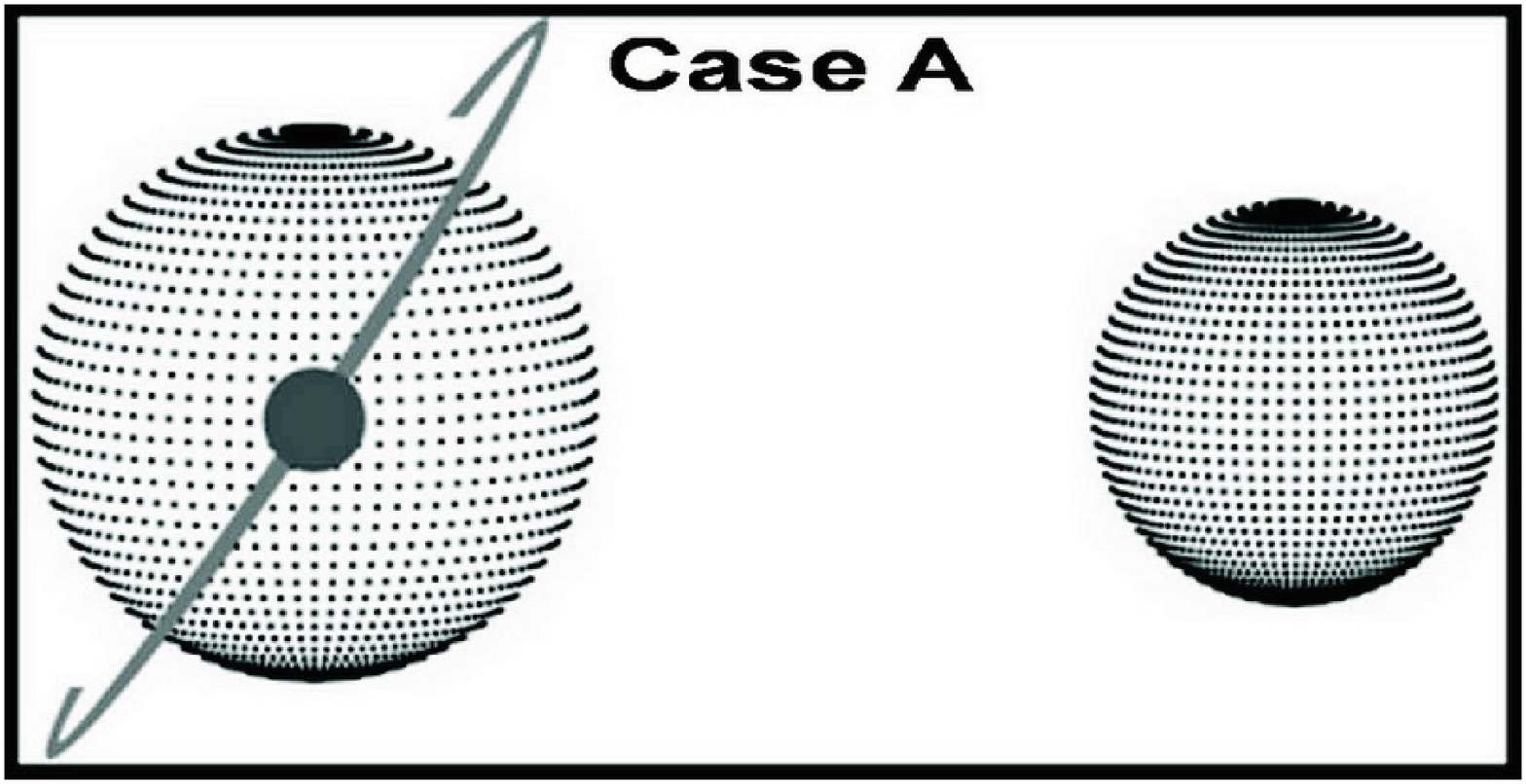}&\includegraphics[width=3.1cm]{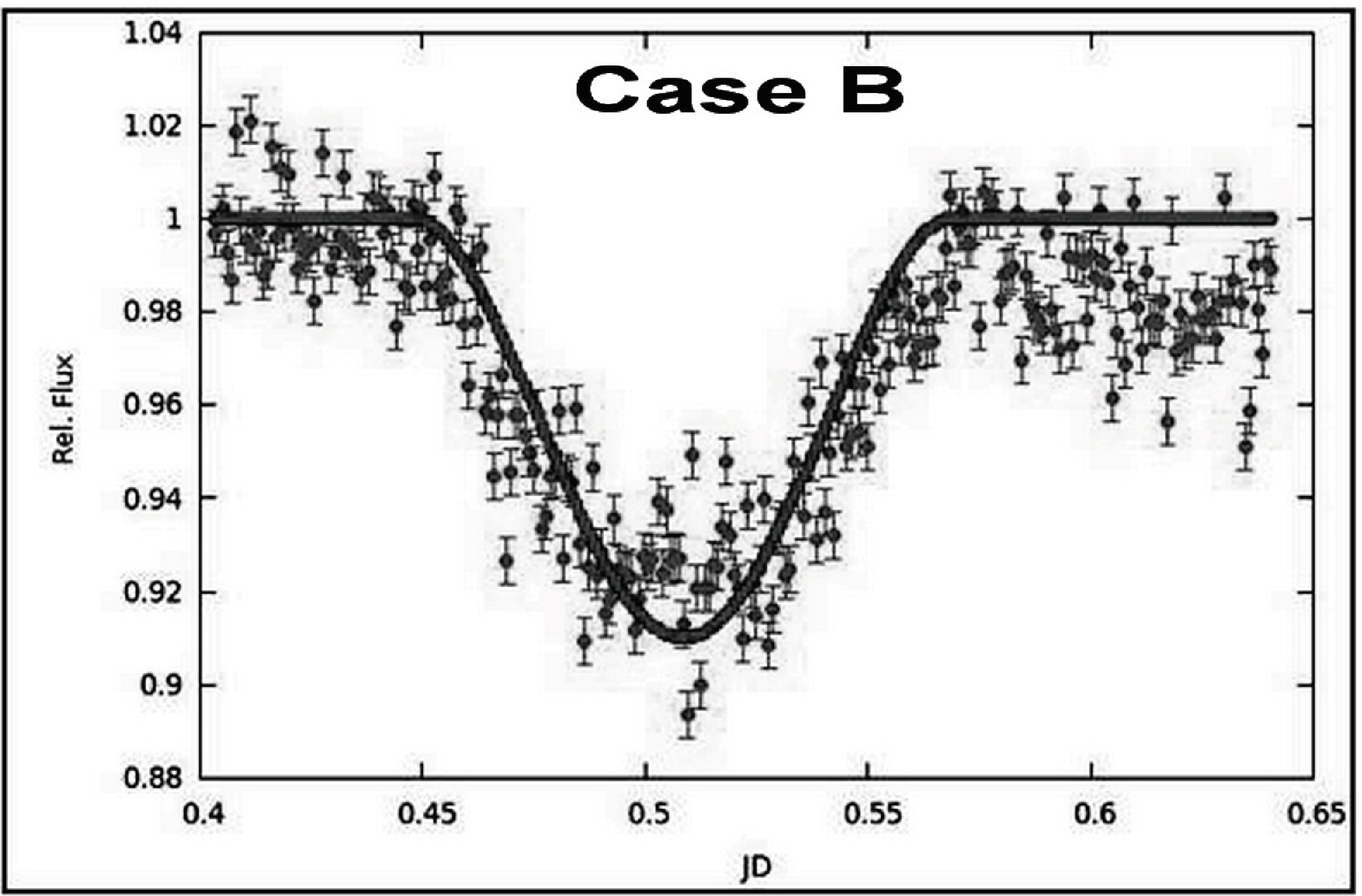}&\includegraphics[width=3.0cm]{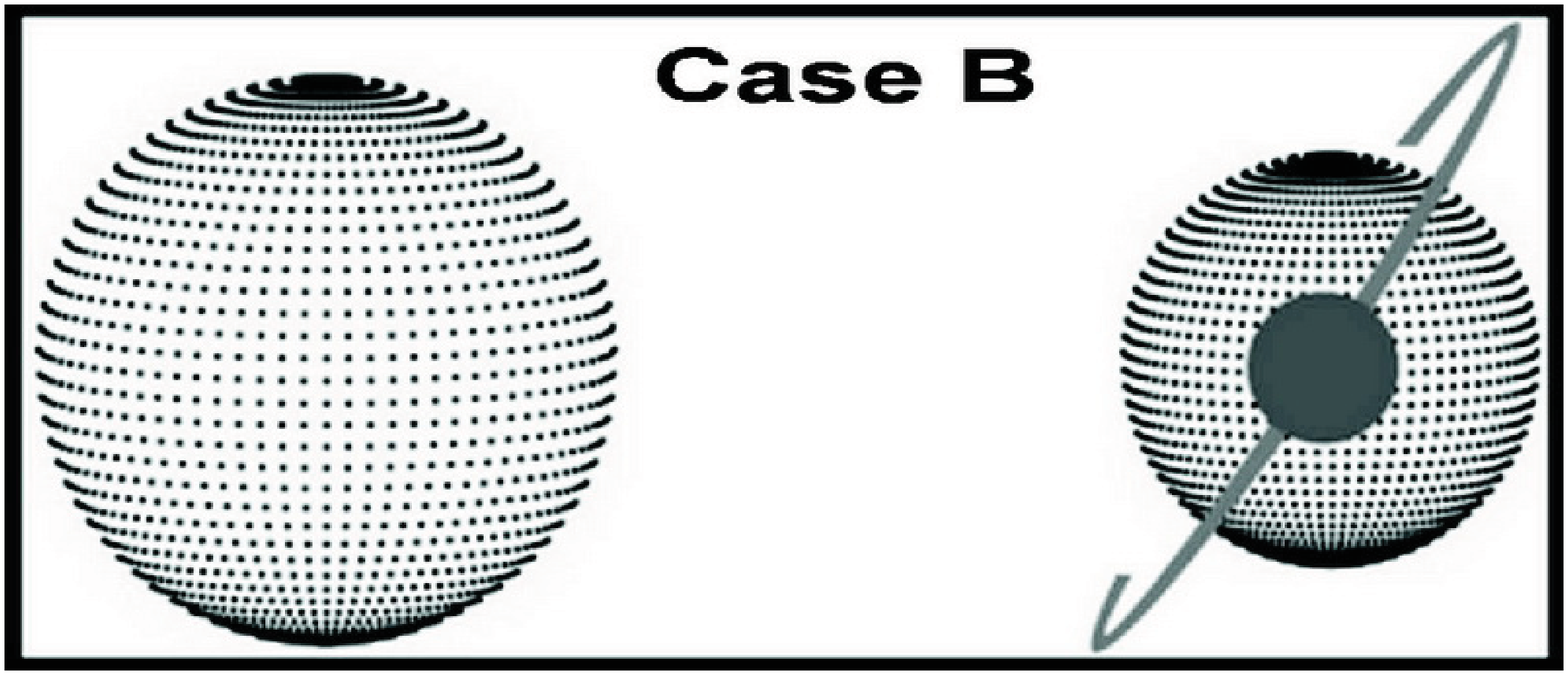}\\
\end{tabular}
\caption{Observed (symbols) and synthetic (solid lines) transit light curve of the third body of AV CMi for each case for HJD 2455538 along with the corresponding 3D plots where the third body and its orbit are shown. }
\label{fig1}
\end{center}
\end{figure}

Five complete transit LCs were analysed with the PhoS-T (\cite[Mislis et al. 2011]{M11}). For a first approach of the third body's parameters, the following hypotheses were adopted: {\bf Case A:} The third body orbits the primary and {\bf Case B:} it orbits the secondary component. For each case the light contribution of the binary component around which the third body is not orbiting (secondary in case A and primary in case B) was subtracted from the total light (of the triple system) by taking into account its fractional luminosity (see Table \,\ref{tab1}) and the residual light curves were re-normalized. The masses of the components and the period of the third body were used to find the semi-major axis of the tertiary component's orbit (assumed a circular one) for each case. The period and the semi-major axis of the third body, the radius and the limb darkening coefficients (\cite[Claret 2004]{C04}) of the `host' component were kept fixed in the programme, while the radius $R_3$ and the inclination $i_3$ of the orbit of the third body were adjusted. A sample of synthetic and observed transit LCs are shown in Fig.\,\ref{fig1}, while the derived parameters from both analyses are listed in Table \,\ref{tab1}.

\begin{table}
\begin{center}
\caption{Light curve and transit analyses results.}
\label{tab1}
\scalebox{0.72}{
\begin{tabular}{ccc cccc ccc c}
\hline
                            \multicolumn{7}{c}{\bf Light curve parameters}                           &     \multicolumn{3}{c}{\bf Absolute parameters} \\
\hline
\emph{Component:}& \emph{Primary}&\emph{Secondary}&\emph{Filters:} &\emph{V} &  \emph{R} &  \emph{I} &\emph{Component:}&\emph{Primary}&\emph{Secondary}\\
\hline
$\Omega$       &   5.73 (1) &    6.76 (1)      &     $x_{1}$   &    0.527    &   0.443   &  0.353    &M [M$_{\odot}$]  &     1.90*    &     1.60 (1)   \\
g              &      1*    &       1*         &     $x_{2}$   &    0.522    &   0.444   &  0.357    &R [R$_{\odot}$]  &    2.38 (5)  &     1.72 (4)   \\
A              &      1*    &       1*         &  L$_1$/L$_T$  &  0.654 (2)  & 0.646 (2) & 0.634 (2) &L [L$_{\odot}$]  &    19.8 (8)  &     10.3 (4)   \\
e              & \multicolumn{2}{c}{0.11 (1)}  &  L$_2$/L$_T$  &  0.341 (1)  & 0.337 (1) & 0.331 (1) &T [K]            &     7900*    &    7897 (8)    \\
i [deg]        & \multicolumn{2}{c}{83.6 (1)}  &  L$_3$/L$_T$  &  0.004 (2)  & 0.016 (2) & 0.036 (3) &a [R$_{\odot}$]  &    5.2 (2)   &     6.2 (1)    \\
q              & \multicolumn{2}{c}{0.843 (3)} &               &             &           &           &log$g$ [cm/s$^2$]&    3.96 (2)  &     4.17 (2)   \\
\hline
                                                                 \multicolumn{11}{c}{\bf Transit parameters}                                           \\
\hline
$HJD$: &\multicolumn{2}{c}{2454521}&   \multicolumn{2}{c}{2454783} &  \multicolumn{2}{c}{2455538}  &  \multicolumn{2}{c}{2455588} &\multicolumn{2}{c}{2455601}\\
\hline
$Case$:&      A       &      B    &       A       &       B       &       A       &       B       &       A       &       B       &       A       &       B  \\
\hline
R$_3$ [R$_{Jup}$]& 4.1 &    5.4    &      4.6      &       6.1     &      4.7      &      6.9      &      4.7      &      6.9      &      4.1      &      6.6 \\
i$_3$ [deg]      & 55.5&    61.9   &     56.1      &      62.3     &      56.8     &      60.0     &      53.7     &     58.1      &     55.6      &     57.7 \\
$\chi^2$     & 2.14    &    9.13   &     2.70      &     7.36      &      2.14     &     9.96      &     1.29      &      5.52     &     1.32      &     6.01 \\       \hline
\multicolumn{11}{l}{*assumed, L$_T$ = L$_1$ + L$_2$ +L$_3$}
\end{tabular}}
\end{center}
\end{table}

\section{Discussion and conclusions}

New LC modelling of AV~CMi and analysis of five transits of the third body in front of one of the components were obtained. The absolute parameters of the eclipsing components were calculated and showed that they are both MS stars with almost the same temperature and eccentric orbits. Using the 18 transit observations an updated ephemeris of the transits was calculated: {\bf T$_{transit}$ = HJD 2454899.354 (1) + 0.519215 (1)$^d \times$E}.
The results of transit analyses showed that both $R_3$ and $i_3$ for each case are varying, probably due to the non-spherical shape of the components. A mean value for the third body's radius yielded as 4.4(3) R$_{Jup}$ and 6.4(6) R$_{Jup}$ for cases A and B, respectively. However, the $\chi^2$ value of case B was found to be systematically greater than that of case A, indicating that the solution of case A is more realistic.
The present results, although they offer a first step for the investigation of the third component in AV~CMi, cannot provide a final conclusion about its nature. The `Hot Jupiter' scenario seems to fail due to the big value of the radius, while, the `Brown dwarf' hypothesis is probably the solution for the nature of the tertiary companion. Moreover, the LC analysis of AV~CMi showed that a third light contribution of $\sim$2\% maybe exists. This value is impossible for planets but not for low-luminosity stars.
High accuracy spectral observations are certainly needed for: (a) spectral classification of the eclipsing components, and (b) radial velocities measurements which will help to derive the mass ratio of the system and probably will reveal the third body's motion, and maybe its nature.



\end{document}